\documentclass[12pt,twoside,fleqn]{article}
\usepackage{epsfig}
\usepackage{psfig}
\def\lsim{\raise0.3ex\hbox{$<$\kern-0.75em\raise-1.1ex\hbox{$\sim$}}}
\def\gsim{\raise0.3ex\hbox{$>$\kern-0.75em\raise-1.1ex\hbox{$\sim$}}}
\makeatletter
\@addtoreset{equation}{section}
\makeatother

\newcommand{\beqn}{\begin{equation}}
\newcommand{\eqn}{\end{equation}}
\newcommand{\bqa}{\begin{eqnarray}}
\newcommand{\eqa}{\end{eqnarray}}
\newcommand{\bqas}{\begin{eqnarray*}}
\newcommand{\eqas}{\end{eqnarray*}}
\newcommand{\bdm}{\begin{displaymath}}
\newcommand{\edm}{\end{displaymath}}

\begin{document}
\thispagestyle{empty}
%
 \mbox{} \hfill BI-TP 2000/20\\
\begin{center}
{{\large \bf The three-dimensional, three-state Potts Model \\
in an External Field} 
 } \\
\vspace*{1.0cm}
{\large Frithjof Karsch and Sven Stickan} 

\vspace*{1.0cm}

{\normalsize
$\mbox{}$ {Fakult\"at f\"ur Physik, Universit\"at Bielefeld,
D-33615 Bielefeld, Germany}
}
\end{center}
\vspace*{1.0cm}
\centerline{\large ABSTRACT}

\baselineskip 20pt

\noindent
We analyze the critical behaviour of the 3-d, 3-state Potts model in the 
presence of an external ordering field. From a finite size 
scaling analysis on lattices of size up to $70^3$  we determine the
critical endpoint of the line of first order phase transitions as
$(\beta_c, h_c) =(0.54938(2), 0.000775(10))$. We determine the relevant 
temperature like and symmetry breaking directions at this second order
critical point and explicitly verify that it is in the universality class 
of the 3-d Ising model.
\vfill

\noindent
July 2000
\eject
\baselineskip 15pt

\section{Introduction}
The temperature driven first order phase transition of the three-dimensional, 
three-state Potts model has been analyzed in great detail in the
absence of a symmetry breaking external field\cite{Blote,Janke}. 
This transition remains first order in the presence of a non-vanishing
external field and for a critical field strength, $h_c$, it ends in a
second order critical point. Although it is expected that this critical
point belongs to the universality class of the 3-d Ising model, the 
universal behaviour in its vicinity has not been analyzed in detail
so far. A first estimate for the location of the critical endpoint
is given in \cite{DeTar1}.

Our interest in properties of the 3-d, 3-state Potts model in the vicinity 
of the second order critical point is motivated by its importance for
the analysis of lines of first order transitions in lattice gauge 
field models. The universal behaviour in the 
vicinity of the second order endpoint of the line of first order
transitions has recently been investigated in the U(1)-Higgs \cite{higgs1} 
and SU(2)-Higgs \cite{higgs} models.  
Lines of first order transitions with critical endpoints do,
however, also occur in QCD with light as well as heavy quarks.  
In the heavy quark mass limit of finite temperature QCD a line of
first order phase transitions (deconfinement transition) occurs
which is closely related to the phase transition in
the 3-d, 3-state Potts model in an external field \cite{Yaffe,Green}.
Like in the Potts model the deconfinement transition is 
first order in the limit of infinitely heavy quarks
and ends in a second order transition at some finite
value of the quark mass. This critical point is expected to belong 
to the universality class of the 3-d Ising model \cite{Pisarski}. 
Another line of first order transitions occurs
in the case of QCD with three light quark flavours (chiral symmetry
restoration). This line also ends in a second order 
endpoint at some critical value of the quark mass. In order to
analyze the critical behaviour at these endpoints it will be important
to disentangle the relevant energy- and ordering field like 
directions and identify the Ising-like observables at the critical points. 
We will address this problem here first in the simpler case of the Potts
model and will explore methods used for the analysis of 
the liquid-gas phase transition \cite{wilding} as well as lattice  
gauge models \cite{higgs1,higgs}.
 
In this letter we present an accurate determination of the critical couplings
at the endpoint of the line of first order phase transitions
in the 3-d, 3-state Potts model with an external
ordering field. Moreover, we will determine the relevant energy-like 
and ordering field like directions at this endpoint as well as the
related operators from which critical exponents and other universal
constants (Binder cumulants) can be extracted. 
In the next section we fix our notation for the Potts model and introduce 
the new couplings and operators which control the critical
behaviour at the endpoint. In section 3 we present our numerical 
results for the location of the critical endpoint. A more detailed
discussion of the universal properties at this endpoint is given in
Section 4. Finally we present our conclusions in Section 5. 

\section{The Model and Simulation Parameters}
The three-state Potts model is described in terms of {\it spin variables}
$\sigma_i \in \{1,~2,~3\}$, which are located at sites $i$ of a cubic
lattice of size $V=L^3$. The Hamiltonian of the model is given by,
\begin{equation}
H= -\beta E - h M \quad,
\label{hamiltonian}
\end{equation}
where $E$ and $M$ denote the energy and magnetization,
\begin{equation}
E=  \sum_{\langle i,j \rangle} \delta(\sigma_i, \sigma_j)\quad , \quad
M=\sum_i \delta(\sigma_i, \sigma_g) \quad .
\label{EandM}
\end{equation}
Here the first sum runs over all nearest neighbour pairs of sites $i$ and $j$.
A non-vanishing field $h > 0$ favours magnetization in the direction of 
the {\it ghost spin} $\sigma_g$. On a finite lattice of size $L^3$ the 
partition function of the model is then given by
\begin{equation}
Z(\beta, h, L) = \sum_{\{ \sigma_i \}} {\rm e}^{-H} \quad .
\label{partition}
\end{equation}
For vanishing external field the model is known to have a first order
phase transition for $\beta_c(h=0) = 0.550565(10)$ \cite{Janke}. 
In the presence of a non-vanishing external field $h$ this first
order transition weakens and ends in a second order critical endpoint,
which is expected to belong to the universality class of the $3$-d Ising
model. A first estimate of the critical endpoint $(\beta_c, h_c)$ has 
been given in Ref.~\cite{DeTar1}. We will give here a more precise 
determination of $(\beta_c, h_c)$ and analyze the universal critical 
behaviour at this point.

At $(\beta_c, h_c)$ the original operators for the energy and 
magnetization, $E$ and $M$, loose their meaning as operators being
conjugate to the temperature-like and symmetry breaking couplings.
One rather has to determine the new relevant directions at the critical
endpoint, which take over the role of temperature-like and symmetry breaking
directions and allow the determination of the two relevant critical 
exponents at the second order endpoint. This also fixes the new order 
parameter and energy-like observables as {\it mixed operators} in terms
of the original variables $E$ and $M$. Following the discussion of the 
corresponding problem for the liquid-gas transition \cite{wilding} we 
introduce new operators 
\begin{equation}
\tilde{M} = M + s E \quad , \quad \tilde{E} = E + r M\quad ,
\label{newEM}
\end{equation}
as superpositions of the original variables $E$ and $M$. The Hamiltonian
of the Potts model can then be rewritten in terms of these new operators,
\begin{equation}
H = -\tau \tilde{E} - \xi \tilde{M} \quad,
\label{newH}
\end{equation}
where the new couplings are given by
\begin{equation}
\xi = {1\over 1-rs}(h-r\beta) \quad , \quad 
\tau = {1\over 1-rs}(\beta -s h) \quad.
\label{newTH}
\end{equation}
We note that the general ansatz given by Eqs.~\ref{newEM} to \ref{newTH}
does allow for the possibility that the new couplings $\tau$ and
$\xi$ do not define orthogonal directions in the space of the original
couplings $\beta$ and $h$, {\it i.e.} they need not result from a
rotation of the couplings $\beta$ and $h$. 

In the presence of a non-vanishing external field the line of first order 
phase transitions, $\beta_c (h)$, singles out a direction which
corresponds to the low temperature, symmetry broken part of a
temperature-driven transition.
The new temperature-like direction $\tau$ thus is identified with 
$\beta_c (h)$.  In the vicinity of the critical endpoint the slope of 
this line determines the mixing parameter $r$,  
\begin{equation}
r^{-1} = \biggl( {{\rm d}\beta_c(h) \over {\rm d}h} \biggr)_{h=h_c} \quad .
\label{mixing}
\end{equation} 
The second mixing parameter, $s$, is determined by demanding that the
energy-like fluctuations and those of the ordering field are uncorrelated,
\begin{equation}
\langle \delta \tilde{M} \cdot \delta \tilde{E} \rangle  \equiv 0
\quad ,
\label{fluctuations}
\end{equation} 
with $\delta \tilde{X} = \tilde{X} - \langle \tilde{X} \rangle$ for
$X=M$ and $E$. 
This insures that the expectation value of the new ordering 
field operator, $\langle \tilde M \rangle$, fulfills a basic
property of an order parameter, {\it i.e.} for $\xi = \xi_c$ it stays 
$\tau$-independent in the symmetric phase. Eqs.~\ref{mixing} and
\ref{fluctuations} give two independent conditions which are sufficient
to determine the mixing parameters $r$ and $s$. 

All our simulations have been performed on lattices of size $L^3$ with
$L=40,~50,~60$ and $70$. We use periodic boundary conditions and perform
simulations with a cluster algorithm described in Ref.~\cite{Wolff}.
Independent of the lattice size we call a {\it new configuration} the
spin configuration obtained after 1000 cluster updates. Typically
we have for each value of the external field $h$ performed simulations 
at 3 to 4 $\beta$-values in the vicinity of the critical point. For each
pair of couplings $(\beta, h)$ we generated about 10000 configurations which 
then have been used in a Ferrenberg-Swendsen reweighting analysis \cite{Fer88}
to calculate observables at intermediate parameter values. Autocorrelation
times have been estimated by monitoring the time evolution of  the energy $E$. 
Close to the pseudo-critical points they vary between 7 and 25 configurations 
on the smallest and largest lattices, respectively. 
All errors quoted below have been obtained from a 
jackknife analysis.

\section{Determination of the critical endpoint}

The basic observables for the determination of the pseudo-critical couplings 
on finite lattices are susceptibilities constructed from the Hamiltonian $H$, 
and the magnetization $M$,
\begin{eqnarray}
c_L &=& {1\over L^{3}} 
\biggl( \langle H^2 \rangle - \langle H \rangle^2 \biggr)\quad ,\\
\chi_L &=& {1\over L^{3}} 
\biggl( \langle M^2 \rangle - \langle M \rangle^2 \biggr)\quad . 
\label{suscept}
\end{eqnarray}   
The location of the maxima in these observables define pseudo-critical 
couplings, $\beta_{c,L}(h)$, which may differ for different
observables. We generally find that the locations of $c_{L,max}$ 
and $\chi_{L,max}$ differ by about one standard deviation of the
statistical errors on $\beta_{c,L}(h)$ for our smaller lattices and 
agree within errors for the largest lattice, $L=70$. 
The maxima have been obtained 
from a combined Ferrenberg-Swendsen analysis which takes into account
all data sets at fixed values of $h$ at $\beta$-values in the vicinity
of the pseudo-critical points. In Tab.~\ref{tab:betac} we quote results 
for $\beta_{c,L}(h)$ obtained from $\chi_{L,max}$ for couplings, $h$, in 
the vicinity of the critical endpoint.
\begin{table}
\begin{center}
\begin{tabular}{|c|c|c|c|c|}\hline  
~&\multicolumn{4}{|c|}{$\beta_{c,L} (h)$}\\
\hline 
$h$&$L=40$&$L=50$&$L=60$&$L=70$\\
\hline 
~0.00050&~0.549800(16)&~0.549798(10)&~0.549792 (8)&~0.549790 (6)\\
~0.00055&~0.549742(16)&~0.549724(12)&~0.549718 (8)&~0.549716 (6)\\
~0.00060&~0.549668(18)&~0.549652(12)&~0.549644 (8)&~0.549642 (6)\\
~0.00065&~0.549596(20)&~0.549578(14)&~0.549572 (8)&~0.549568 (6)\\
~0.00070&~0.549524(20)&~0.549506(14)&~0.549498 (8)&~0.549496 (6)\\
~0.00075&~0.549454(20)&~0.549434(14)&~0.549426 (8)&~0.549424 (6)\\
~0.00080&~0.549382(22)&~0.549362(16)&~0.549354(10)&~0.549352 (6)\\
~0.00085&~0.549312(22)&~0.549292(16)&~0.549282(10)&~0.549280 (8)\\
~0.00090&~0.549240(22)&~0.549220(18)&~0.549212(10)&~0.549208(10)\\
~0.00095&~0.549170(24)&~0.549150(18)&~0.549140(10)&~0.549136(10)\\
\hline 
\end{tabular}
\end{center}
\caption{Pseudo-critical couplings determined from the location of
the peak in $\chi_L$ on different size lattices at 
various values of the external field $h$.}
\label{tab:betac}
\end{table}

\begin{figure}
\label{fig:chipeak}
\begin{center}
\epsfig{file=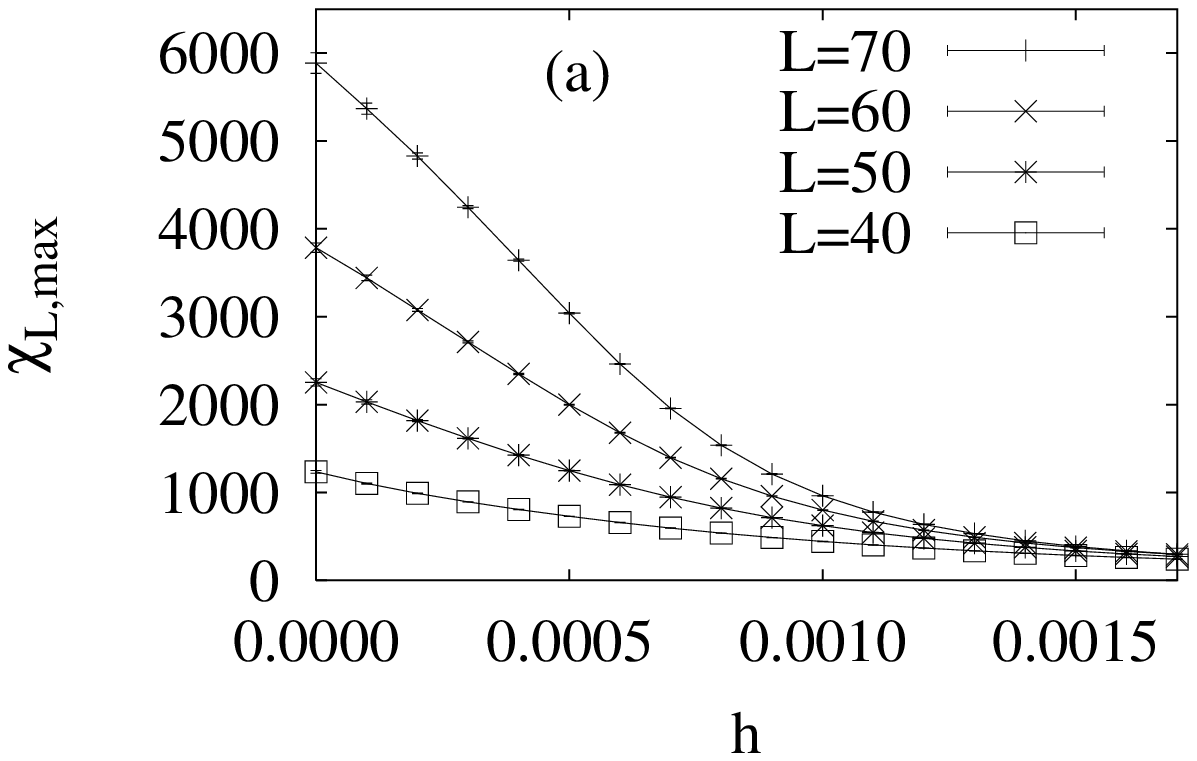,width=100mm}
\epsfig{file=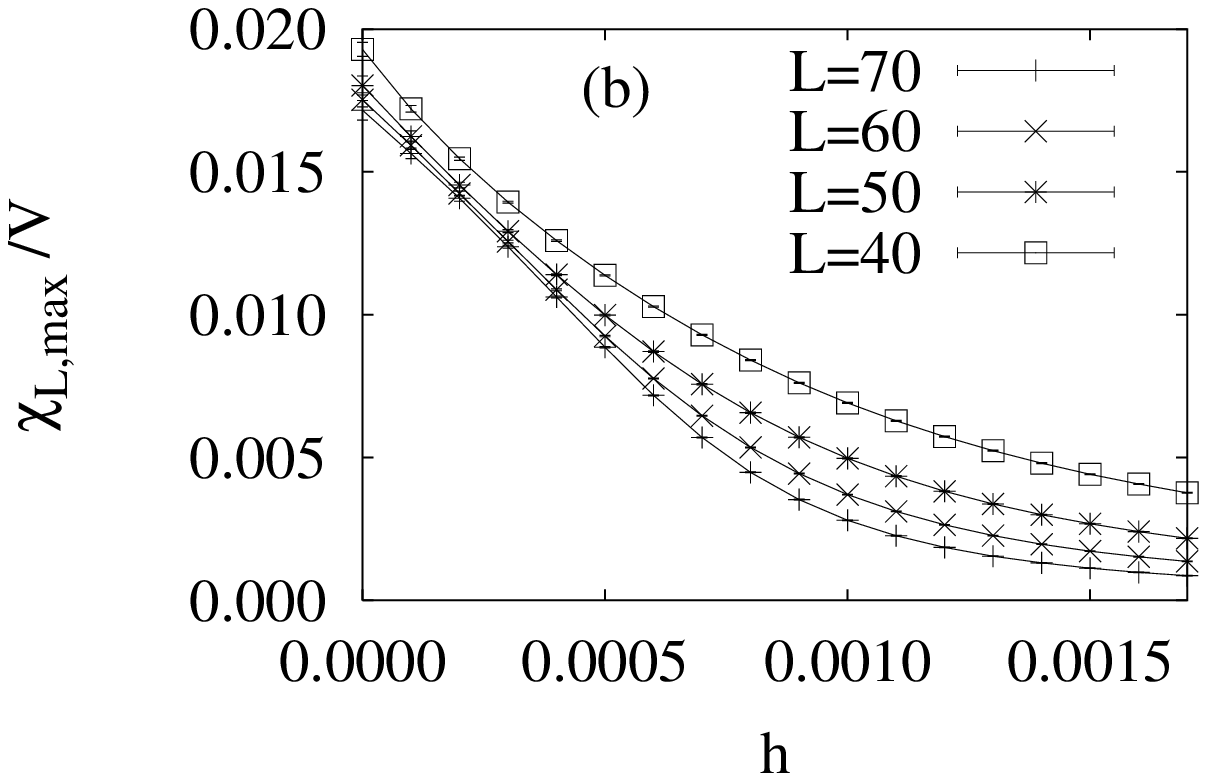,width=100mm}
\end{center}
\caption{The maxima of $\chi_L$ (a) and $\chi_L$ in units of the 
volume $V=L^3$ (b) as a function of the external field $h$ on various 
size lattices.} 
\end{figure}

A first indication for the location of the critical region is obtained
from an analysis of the volume dependence of the peak heights in
the susceptibilities. In the region of first order phase transitions 
($h<h_c$) $\chi_{L,max}$ is expected to increase proportional to the
volume, while for $h>h_c$ the peak heights will approach a finite 
value in the infinite volume limit. Results for $\chi_{L,max}$ obtained
on different size lattices are shown in Fig.~1. From this figure as well 
as from a similar analysis
of $c_{L,max}$ it is clear that the critical endpoint will be located
in the interval $0.0005 \le h \le 0.001$. In this interval we find
that the dependence of the pseudo-critical couplings $\beta_{c,L}(h)$  on 
the external field $h$ is well approximated by a leading order 
Taylor-expansion,
\begin{equation}
\beta_{c,L} (h_1) - \beta_{c,L} (h_2) = {1 \over r_L } \bigl( h_1 - h_2 \bigr) 
\quad , \quad h_1, h_2 \in [0.0005, 0.001] \quad .
\label{slope}
\end{equation} 
For our largest lattice, $L=70$, this dependence is shown in 
Fig.~2. A straight line fit to these data yields, 
$r_{70} =-0.689(8)$. 
Results from other lattice sizes are summarized in Tab.~\ref{tab:slope}.
The slope parameter $r_L$ is slightly volume dependent which, of course, 
reflects the volume dependence of the pseudo-critical couplings. We thus
have extrapolated $r_L$ to the infinite volume limit using the ansatz
$r_L=r_{\infty}+ c/L^3$. This yields $r_{\infty}=-0.685(10)$ and fixes  
the mixing parameter $r$ defined in Eq.~\ref{newEM}. 

\begin{figure}
\label{fig:slope}
\begin{center}
\epsfig{file=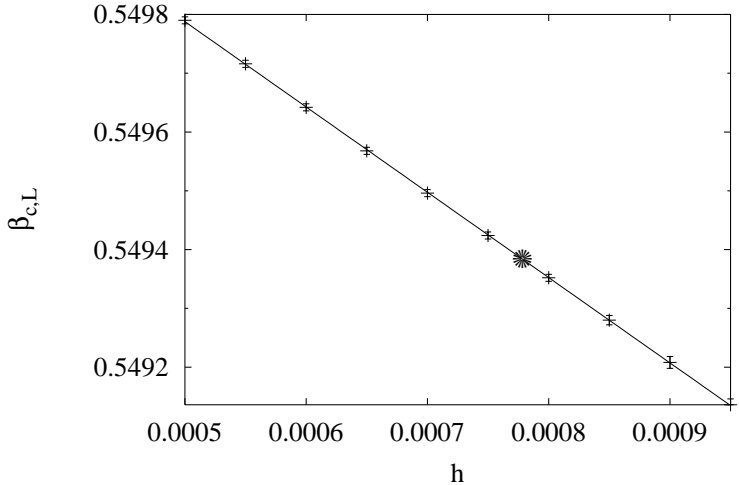,width=100mm}
\end{center}
\caption{The dependence of the critical coupling on the external field
on a $70^3$ lattice in the vicinity of the second order endpoint whose
location is indicated by an asterisk.} 
\end{figure}

\begin{table}
\begin{center}
\begin{tabular}{|c|c|c|c|c|c|}\hline  
$L$&40&50&60&70&$\infty$\\
\hline 
$r_L$&-0.706(21)&-0.694(15)&-0.690 (9)&-0.689 (8)&-0.685(10)\\
$s_L$& 0.696 (1)& 0.696 (1)& 0.694 (1)& 0.690 (3)&0.690(2)\\
\hline 
\end{tabular}
\end{center}
\caption{Slope parameter, $r_L$, determined from the dependence of the 
pseudo-critical couplings on the external field $h$ (Eq.~\ref{mixing}) and 
the second mixing parameter $s_L$ determined from Eq.~\ref{fluctuations}
for different lattice sizes.}
\label{tab:slope}
\end{table}

The mixing parameter $r_L$ determined above may be used in connection
with Eq.~\ref{fluctuations} to determine the second mixing parameter $s_L$.
We have done so and within statistical errors we found $s_L=-r_L$, {\it i.e.}
the couplings $(\tau,\xi)$ can be obtained from a rotation in the space
of the original couplings $(\beta, h)$.  An alternative approach thus is to 
assume $s_L=-r_L$ and use Eq.~\ref{fluctuations} to determine $s_L$. 
In this way we also can make use of the information on diagonal
correlations $\langle (\delta X)^2 \rangle$, $X=E,~M$ and determine
$s_L$ from a diagonalization of the fluctuation matrix
\cite{higgs1,higgs}. 
The results obtained in this way for $s_L$ on different size lattice 
are also given in Tab.~\ref{tab:slope}. We note that within errors this  
approach is consistent with the determination of $r_L$ from the slope
of the critical line. The statistical errors are, however, significantly
smaller. This shows that the transformation of variables, 
$(\beta,h) \rightarrow (\tau,\xi)$ defined in Eq.~\ref{newTH} indeed 
is a rotation.
In the following we will use for the mixing parameters $r=-s$ with 
$s=0.690$.

\begin{figure}
\label{fig:binder}
\begin{center}
\epsfig{file=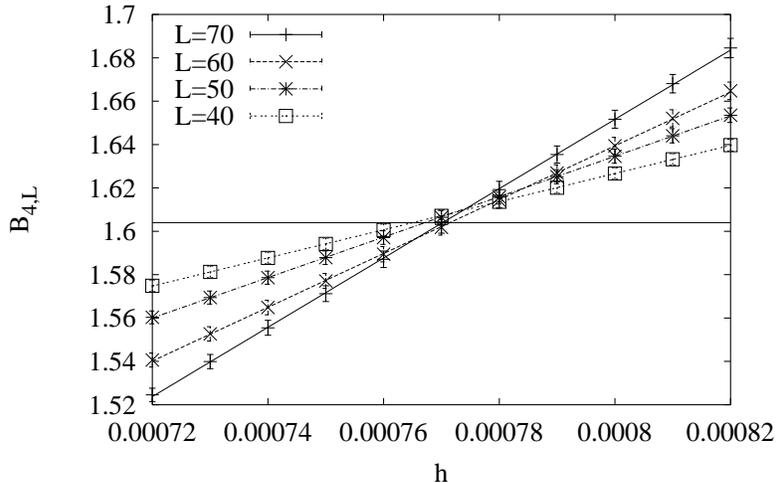,width=100mm}
\end{center}
\caption{The Binder cumulant $B_{4,L}$ versus $h$ for different 
lattice sizes. The horizontal line shows the universal value of
the Binder cumulant for the universality class of the three dimensional
Ising model, $B_4 = 1.604(1)$ \cite{Blote95}.} 
\end{figure}

Having fixed the mixing parameters we can perform an
analysis of the critical behaviour in the vicinity of the critical 
endpoint using standard techniques to study temperature driven second order 
phase transitions in the absence of an external 
symmetry breaking field. In 
particular, we can use Binder cumulants to locate the critical endpoint,
\begin{equation}
B_{3,L} = {\langle (\delta \tilde{M})^3 \rangle \over \langle 
(\delta \tilde{M})^2 \rangle^{3/2} }
\quad , \quad 
B_{4,L} = {\langle (\delta \tilde{M})^4 \rangle \over \langle 
(\delta \tilde{M})^2 \rangle^2 }\quad .
\label{binder}
\end{equation}
For given values of $h$ and $L$ we find that the cumulant $B_{3,L}(\beta,h)$
vanishes and $B_{4,L}(\beta,h)$ acquires a minimum at values of the coupling 
$\beta$ which agree with each other within statistical errors. This, in turn, 
defines a pseudo-critical coupling, which again agrees within errors with the 
pseudo-critical couplings given in Tab.~\ref{tab:betac}.
The minima of the second Binder cumulant, $B_{4,L}$, calculated on 
different size lattices should have a unique crossing point when plotted,
for instance, versus the external field $h$. This is indeed the case,
as can be seen in Fig.~\ref{fig:binder}. The crossing point yields
the critical field $h_c$ at the second order endpoint.

From the crossing points of the Binder cumulants we find for the 
critical endpoint
\begin{equation} 
\beta_c = 0.54938(2) \quad , \quad 
h_c = 0.000775(10) \quad .
\label{endpoint}
\end{equation}
or equivalently in terms of the rotated couplings, 
\begin{equation} 
(\tau_c,\xi_c)=(0.37182(2),0.25733(2)) \quad.
\label{endpointr}
\end{equation}
At this point the Binder cumulant takes on the value $B_4 = 1.609(4)(10)$, 
where the first error only takes into account the fluctuation of $B_{4,L}$
on the different size lattices at $(\beta_c,h_c)$ and the second error
estimates the uncertainties arising from the errors on the location of
the endpoint. Our result agrees within errors with the value found for the 
three dimensional Ising model, $B_4 = 1.604(1)$ \cite{Blote95}.

\section{Universality class of the endpoint}

\begin{figure}
\label{fig:histo}
\begin{center}
\hspace{-22mm}\epsfig{file=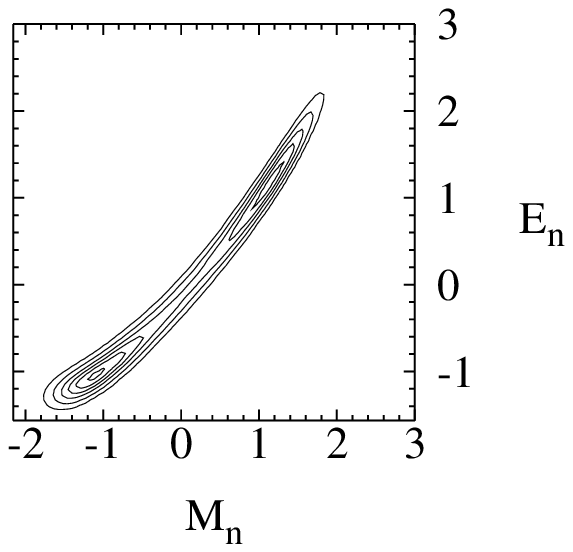,width=95mm}
\hspace{-20mm}\epsfig{file=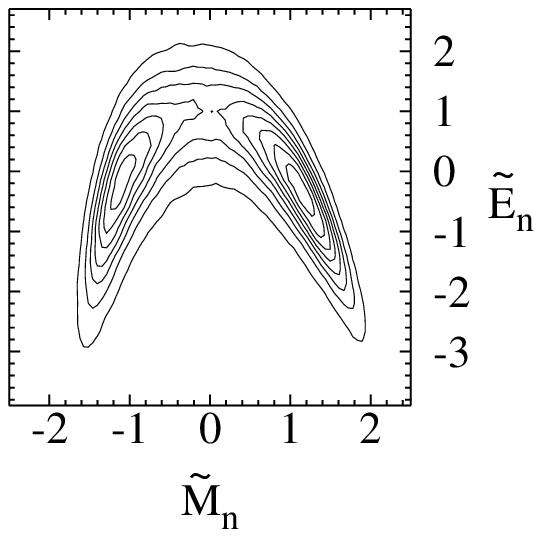,width=95mm}
\end{center}
\vspace{-2.0cm}
\caption{Correlations between the original fields $E$ and $M$ (left)
and the new energy-like and ordering field like operators $\tilde E$ and
$\tilde M$. Shown are contour plots obtained from joint histograms for
these observables on a $70^3$ lattice at $(\beta_c,h_c)$. The
histograms have been obtained from a Ferrenberg-Swendsen reweighting
in $\beta$ and $h$. Actually shown are the normalized observables
which have vanishing expectation value and a variance that is equal
to unity,
$(X-\langle X \rangle)/\langle (X-\langle X \rangle)^2\rangle^{1/2}$. }
\end{figure}

The value of the Binder cumulant $B_4$ determined at the critical
endpoint strongly suggests that the critical endpoint belongs to the 
universality class of the 3-d Ising model.
Rather impressive support for Ising-like behaviour at the critical
endpoint, which at the same time also demonstrates the importance
of the correct choice of energy- and ordering field like variables,
is given by the contour plots obtained from the joined 
probability distributions (two-dimensional histograms)  
of the operators $(E,M)$ and $(\tilde E , \tilde M )$, respectively.
These are shown in Fig.~\ref{fig:histo} for the  $70^3$ lattice at
$(\beta_c,h_c)$. Although the contour plot for $(\tilde E , \tilde M )$
is still slightly asymmetric, a comparison with the corresponding contour 
plot of the 3-d Ising model \cite{higgs} clearly shows the same 
characteristic {\it fingerprint}. We have checked that this asymmetry 
is well within the uncertainties of our determinations of $(\beta_c,h_c)$ 
as well as the mixing parameters $r$ and $s$. 

Further support for the Ising universality class of the critical
endpoint comes from a more conventional finite size scaling analysis  
performed for the
newly defined mixed operators $\tilde M$, $\tilde E$ and the related
susceptibilities. In particular we have considered the order parameter 
\begin{equation}
\tilde{m}(\tau,\xi) = {1\over L^3} \biggl( \tilde{M}(\tau,\xi) -\langle
\tilde{M}(\tau_c,\xi_c)\rangle \biggr) \quad ,
\label{neworder}
\end{equation}
and the susceptibility 
\begin{equation}
\tilde{\chi}_L = L^3 \biggl( \langle \tilde{m}^2 \rangle - 
\langle |\tilde{m}| \rangle^2 \biggr)\quad . 
\label{newchi}
\end{equation}
These observables indeed show the usual finite size behaviour in the
vicinity of the critical point. For fixed $\xi\equiv \xi_c$ the 
susceptibility $\tilde{\chi}_L$ reaches a maximum at a pseudo-critical 
coupling $\tau_{pc}$. Critical exponents can then be determined from 
the scaling of the order parameter, 
$\langle |\tilde{m}| \rangle \sim L^{-\beta/\nu}$, the peak height of
the susceptibility, 
$\tilde{\chi}_{L,max} \sim L^{-\gamma/\nu}$, and the 
pseudo-critical couplings, $(\tau_{pc} - \tau_c) \sim L^{-1/\nu}$.
In our analysis of the critical behaviour we have first fixed the critical
point $(\tau_c,\xi_c)$ and determined critical exponents from 
two-parameter fits. Errors have then be determined by also varying
$(\tau_c,\xi_c)$ within the errors given in Eq.~\ref{endpointr}. 
In Tab.~\ref{tab:exponents} we summarize the results of this analysis 
and compare the calculated exponents with known values for the 3-d 
Ising model.  As can be seen the agreement is quite satisfactory
although in our analysis the lattices have not yet been large enough to
reach an accuracy similar to that obtained for the 3-d Ising model
\cite{Janke,Blote95}.    

\begin{table}
\begin{center}
\begin{tabular}{|c|c|c|}\hline  
$~$&this work&3-d Ising\\
\hline 
$1/\nu$&1.60(2)&1.587(2)\\
$\beta/\nu$&0.517(3)&0.5185(15)\\
$\gamma/\nu$&1.93(1)&1.9630(30)\\
$(1-\alpha)/\nu$&1.30(2)&1.413(2)\\
\hline 
\end{tabular}
\end{center}
\caption{Critical exponents determined from a finite size scaling
analysis in the vicinity of the critical endpoint of the Potts model.
The last column is based on the exponents for the 3-d Ising model 
given in \cite{Blote95}.}
\label{tab:exponents}
\end{table}

Most sensitive to the correct choice of the new, mixed observables
are the new energy-like observable,
$\langle \tilde e \rangle = \langle \tilde{E}/L^3\rangle,$
and the corresponding susceptibility, 
$c_e \sim (\partial \tilde e /\partial \tau)_\xi$. Their critical behaviour is
controlled by the thermal critical exponent ($y_t$) which fixes the
critical exponent $\alpha$. In particular, the energy density calculated
at $(\tau_c,\xi_c)$ is  expected to scale like 
$\langle \tilde e \rangle = c_0 + c_1 L^{-(1-\alpha)/\nu}$ at $(\tau_c,\xi_c)$. 
It is obvious from Eq.~\ref{newEM} that this scaling behaviour can
hold only if in the mixed observable $\tilde e$ the leading singular
behaviour of $E/L^3$ and $M/L^3$, which in general is proportional to
$L^{-\beta/\nu}$, gets canceled.  
For the 3-d Ising model the exponent $\alpha$ is small ($\alpha =
0.11$).
As a consequence the dominant singular behaviour in 
$\tilde e$ is expected to behave like $\sim L^{-1.41}$ rather than
$\sim L^{-0.52}$ as would be the case for $E/L^3$ and $M/L^3$ separately.
As can be seen from Tab.~\ref{tab:exponents} this is supported by our 
analysis of the finite size scaling behaviour of the energy-like observable 
$\langle \tilde{e} \rangle$, although in this case the critical exponent 
shows the largest deviation from the corresponding value of the 3-d Ising 
model. 

\section{Conclusions}

We have determined the critical endpoint of the 3-d, 3-state Potts model
in the presence of an external ordering field and have analyzed
the critical behaviour in the vicinity of this second order phase
transition. New energy-like and ordering field like 
couplings and observables have been obtained through a rotation of the
original couplings and fields. At the critical endpoint of the line of
first order phase transitions the joined probability distribution of
these new observables shows a correlation pattern which is
characteristic for the 3-d Ising model. A finite size scaling analysis
performed with the newly defined rotated observables further supports 
that the critical endpoint belongs to the universality class of the 3-d 
Ising model.

\vspace{0.5cm}
\noindent
{\bf Acknowledgements:} 

\medskip
\noindent
We thank J\"urgen Engels and Andreas Peikert for helpful discussions.
This work has been supported by the TMR network ERBFMRX-CT-970122 and the 
DFG under grant Ka 1198/4-1. 


\end{document}